
\input harvmac
\input epsf
\def\lsim{\mathrel{\rlap{\lower4pt\hbox{\hskip1pt$\sim$}}
    \raise1pt\hbox{$<$}}}         
\noblackbox
\pageno=0\nopagenumbers\tolerance=10000\hfuzz=5pt
\baselineskip=12pt plus2pt minus2pt
\line{\hfill\tt hep-ph/9510449}
\line{\hfill CERN-TH/95-266}
\line{\hfill Edinburgh 95/556}
\line{\hfill GeF-TH-9/95}
\vskip 8pt
\centerline{\bf NEXT-TO-LEADING ORDER DETERMINATION OF THE SINGLET AXIAL}
\centerline{\bf CHARGE AND THE POLARIZED GLUON CONTENT OF THE NUCLEON}
\vskip 12pt\centerline{Richard~D.~Ball\footnote{\dag}{Supported in
part by a Royal Society University Research Fellowship.}}
\vskip 4pt
\centerline{\it Theory Division, CERN,}
\centerline{\it CH-1211 Gen\`eve 23, Switzerland.}
\centerline{\it and}
\centerline{\it Department of Physics and Astronomy}
\centerline{\it University of Edinburgh, Edinburgh EH9 3JZ, Scotland}
\vskip 8pt
\centerline{Stefano~Forte\footnote{\ddag}{On leave
 from INFN, Sezione di Torino, via P.~Giuria 1,
I-10125 Turin, Italy (address after December 1, 1995).}}
\vskip 4pt
\centerline{\it Theory Division, CERN,}
\centerline{\it CH-1211 Gen\`eve 23, Switzerland.}
\vskip 8pt
\centerline{Giovanni Ridolfi}
\vskip 4pt
\centerline{\it INFN,
Sezione di Genova}\centerline{\it Via Dodecaneso 33, I-16146, Genova, Italy}
\vskip 24pt
{\centerline{\bf Abstract}
\medskip\narrower
\baselineskip=10pt
We perform a full next-to-leading analysis of the the available
experimental data on the polarized structure function $g_1$ of the
nucleon, and give a precise determination of its singlet axial charge
together with a thorough assessment of the theoretical uncertainties.
We find that the data are now sufficient to separately determine
first moments of the polarized quark and gluon distributions, and
show in particular that the gluon contribution is large
and positive.
\smallskip }
\vskip 24pt
\centerline{Submitted to: {\it Physics Letters B}}
\vskip 48pt
\line{CERN-TH/95-266\hfill}
\line{October 1995\hfill}

\vfill\eject \footline={\hss\tenrm\folio\hss}

\def\as{\alpha_s}
\def\a{{\as\over2\pi}}

\def\neath#1#2{\mathrel{\mathop{#1}\limits_{#2}}}

\def\epm#1#2{\hbox{${+#1}\atop {-#2}$}}
\def\neath#1#2{\mathrel{\mathop{#1}\limits_{#2}}}
\def\gsim{\mathrel{\rlap{\lower4pt\hbox{\hskip1pt$\sim$}}
    \raise1pt\hbox{$>$}}}         

\def\frac#1#2{{{#1}\over {#2}}}
\def\half{\hbox{${1\over 2}$}}
\def\1{\;1\!\!\!\! 1\;}

\def\smallfrac#1#2{\hbox{${{#1}\over {#2}}$}}

\def\MS{\hbox{$\overline{\rm MS}$}}

\catcode`@=11 
\def\slash#1{\mathord{\mathpalette\c@ncel#1}}
 \def\c@ncel#1#2{\ooalign{$\hfil#1\mkern1mu/\hfil$\crcr$#1#2$}}
\def\lsim{\mathrel{\mathpalette\@versim<}}
\def\gsim{\mathrel{\mathpalette\@versim>}}
 \def\@versim#1#2{\lower0.2ex\vbox{\baselineskip\z@skip\lineskip\z@skip
       \lineskiplimit\z@\ialign{$\m@th#1\hfil##$\crcr#2\crcr\sim\crcr}}}
\catcode`@=12 

\def\PR{{\it Phys.~Rev.~}}
\def\PRL{{\it Phys.~Rev.~Lett.~}}
\def\NP{{\it Nucl.~Phys.~}}
\def\NPBPS{{\it Nucl.~Phys.~B (Proc.~Suppl.)~}}
\def\PL{{\it Phys.~Lett.~}}

\def\ZP{{\it Zeit.~Phys.~}}

\def\vol#1{{\bf #1}}\def\vyp#1#2#3{\vol{#1} (#2) #3}

The polarized structure function $g_1(x,Q^2)$ of the nucleon has been
recently
measured with good accuracy for proton~\ref\proex{E143 Collaboration,
K.~Abe et al., \PRL\vyp{74}{1995}{346}\semi SMC Collaboration,
D.~Adams et al.,
\PL\vyp{B329}{1994}{399}.} and deuteron~\ref\deuex{E143 Collaboration,
K.~Abe et al., \PRL\vyp{75}{1995}{25}\semi SMC Collaboration, D.~Adams et al.,
\PL\vyp{B357}{1995}{248}.}  targets over a reasonably wide range of values of
$x$. This opens up the possibility of a precise determination of the
first moments of $g_1$, which are of direct physical interest, being
related to the nucleon matrix elements of axial currents.
An extraction of the moments of $g_1$
{}~from the data, however, requires theoretical input, not only because the
data
cover a limited range in $x$ but, more importantly, because data are
obtained at
different values of $Q^2$ for each $x$ bin, and have thus to be evolved to a
common value of $Q^2$ using the Altarelli-Parisi equations before the moments
can be evaluated (for recent reviews on the phenomenology of polarized
structure
functions see ref.~\ref\revs{G.~Altarelli and G.~Ridolfi,
\NPBPS\vyp{39B}{1995}{106}\semi S.~Forte, {\tt hep-ph/9409416}, in ``Radiative
Corrections: Status and Outlook'', B.~F.~L.~Ward, ed. (World Scientific,
Singapore, 1995)\semi
R.~D.~Ball, in the proceedings of the ``International School
on Nucleon Spin Structure'', Erice, August 1995 (to be published).}).
We have recently shown~\ref\BFR{R.~D.~Ball, S.~Forte and
G.~Ridolfi, {\it Nucl. Phys.} {\bf B 444} (1995) 287.} that these perturbative
evolution effects can actually be quite large and substantially affect the
extraction of the moments from $g_1$. Furthermore, effects which
are formally next-to-leading order (NLO) may lead to significant evolution
because of the large contribution of the polarized gluons to $g_1$
driven by the axial anomaly.

The recent computation of the full matrix of two-loop anomalous
dimensions~\ref\merner{R.~Mertig and W.~L.~van~Neerven, Leiden preprint
INLO-PUB-6/95; Erratum, private communication (November 1995).}
makes a consistent NLO analysis of $g_1$
now possible. It is the purpose of this paper to perform such an analysis, and
use it to provide a precise determination of the first moment of
$g_1$. After reviewing the
NLO formalism, discussing scheme dependence, and the effect of NLO
corrections on the LO small-$x$ behaviour of parton distributions, we
will use it to extract polarized parton distributions from the data.
We will find that the data allow a determination
of both the quark and gluon distributions without the need for extra
theoretical assumptions, and in particular strongly constrain their
overall normalizations and small-$x$ behaviours: we will
thus be able to infer the existence of
polarized gluons in the nucleon from an analysis of the scale
dependence of $g_1$.
 We will then use
these parton distributions to determine the
first moment of the structure function $g_1$ and  the nucleon
matrix element of the singlet axial current, or singlet axial charge, whose
unexpected smallness has attracted a good deal of interest. We
will finally provide an evaluation of the various sources of
statistical and systematic error related to these determinations.

The structure function $g_1$ is related to the polarized quark and gluon
distributions by
\eqn\gone
{g_1(x,Q^2)=
\smallfrac{\langle e^2\rangle}{2}\left[
C_{\rm NS}\otimes\Delta q_{\rm NS}
+ C_{\rm S}\otimes \Delta\Sigma
+ 2n_f C_g\otimes\Delta g\right],}
where $\langle e^2\rangle=\smallfrac{1}{n_f}\sum_{i=1}^n e^2_i$,
$\otimes$ denotes the usual convolution with respect to
$x$, the nonsinglet and singlet quark distributions are defined as
\eqn\qppd{
\Delta q_{\rm NS}=\sum_{i=1}^{n_f}
\left(\smallfrac{e_i^2}{\langle e^2\rangle}-1\right)
(\Delta q_i+\Delta\bar q_i),\qquad
\Delta \Sigma =\sum_{i=1}^{n_f} (\Delta q_i+\Delta\bar q_i),}
where $\Delta q_i$ and $\Delta\bar q_i$ are the polarized quark
and antiquark distributions of flavor $i$, and $\Delta g$ is the
polarized gluon distribution (see ref.~\BFR\ for further details
of notations and conventions.).
The polarized parton distributions evolve according to the Altarelli-Parisi
equations~\ref\ap{
G.~Altarelli and G.~Parisi, \NP\vyp{B126}{1977}{298}.}
\eqn\aps{\eqalign{\frac{d}{dt}\Delta q_{\rm NS}
&=\frac{\as(t)}{2\pi} P^{\rm NS}_{qq}\otimes\Delta q_{\rm NS}\cr
\frac{d}{dt}\pmatrix{\Delta\Sigma\cr \Delta g\cr}&=
\frac{\as(t)}{2\pi}
\pmatrix{&P^{\rm S}_{qq}& 2 n_f
P^{\rm S}_{qg}\cr &P^{\rm S}_{gq}&
P^{\rm S}_{gg}\cr}
\otimes \pmatrix{\Delta\Sigma\cr \Delta g\cr},\cr}}
where $t\equiv\ln(Q^2/\Lambda^2)$.
The coefficient functions $C$ and splitting functions $P$ may each
be expanded in powers of $\alpha_s$: at NLO
\eqnn\cf\eqnn\splf
$$\eqalignno{C(x,\alpha_s)&=C^{(0)}(x)+
\a C^{(1)}(x)+O(\alpha_s^2)&\cf\cr
P(x,\alpha_s)&=P^{(0)}(x)+
{\as\over2\pi} P^{(1)}(x)+O(\alpha_s^2).&\splf\cr}$$
In accordance with the partonic picture
$C_{\rm NS}^{(0)}(x)=C_{\rm S}^{(0)}(x)=\delta(1-x)$, while
$C_g^{(0)}(x)=0$. It will also prove convenient to introduce
anomalous dimensions
$\gamma(N,\alpha_s)\equiv
\int_0^1\!dx\, x^{N-1} P(x,\alpha_s)$, i.e. the Mellin
transforms
of the splitting functions, as well as analogously defined
moment-space coefficient functions
$C(N,\alpha_s)$ and parton distributions
$\Delta q_{\rm NS}(N,Q^2)$,
$\Delta \Sigma(N,Q^2)$ and $\Delta g(N,Q^2)$.

Whereas the NLO coefficient functions $C^{(1)}$ have been known for
some time~\ref\koda{J.~Kodaira et al., \PR\vyp{D20}{1979}{627};
\NP\vyp{B159}{1979}{99}\semi J.~Kodaira, \NP\vyp{B165}{1979}{129}.} the
two loop splitting functions $P^{(1)}$ have been only recently
determined~\merner,\foot{With our conventions the expressions of $\gamma^{(0)}$
and $\gamma^{(1)}$ of ref.~\merner\ must be divided by $-4$ and $-8$
respectively and $\gamma_{qg}$ should be further divided by $2 n_f$.
} only their first moments being known previously~\koda. The
NLO coefficient functions may be modified by a change of factorization scheme
which is partially compensated by a corresponding change in the NLO
anomalous dimensions, hence both are required
in order to specify a NLO computation completely. Previous analyses which
included NLO effects only in the coefficient functions (such as~\BFR) were thus
necessarily incomplete and treated only the first moments consistently at NLO.
It is now possible to test explicitly whether, as was claimed in
ref.~\BFR, the NLO gluon contribution to the first moment of $g_1$ is
the dominant NLO effect, and furthermore whether it has a sizable
effect on the $Q^2$ dependence of $g_1$.

The NLO anomalous dimensions and coefficient functions of ref.~\merner\
are given in the \MS\ scheme. Since chiral symmetry is respected,
matrix elements of nonsinglet axial currents are conserved,
nonsinglet axial charges do not evolve, and thus $\Delta q_{\rm NS}(1,Q^2)$ is
independent of $Q^2$. In all such schemes at NLO
\eqn\qucoup
{C_{\rm NS}(1,\alpha_s) =C_{\rm S}(1,\alpha_s) =1-\frac{\as}{\pi}
+O(\alpha_s^2)}
(although at higher orders $C_{\rm NS}(1,\alpha_s) \not
=C_{\rm S}(1,\alpha_s)$).  Matrix elements of the axial singlet current
 are instead not conserved
because of the axial anomaly, so that the singlet axial charge
depends on scale. In  the \MS\ scheme the first moment of $C_g^{(1)}$
vanishes, the gluon decouples from the first moment of $g_1$
and the scale dependent singlet axial charge is thus equal to
$\Delta \Sigma (1,Q^2)$.

Factorization schemes where this happens are somewhat pathological, in
that they include soft contributions to
the cross section in the coefficient function rather than absorbing
them completely into the parton distributions~\nref\ccm{R.~D.~Carlitz,
J.~C.~Collins and A.~H.~Mueller,\PL\vyp{B214}{1988}{229}.}\nref\alam{
G.~Altarelli and B.~Lampe, \ZP\vyp{C47}{1990}{315}.}\nref\vogel{
W.~Vogelsang, \ZP\vyp{C50}{1991}{275}.}\refs{\ccm-\vogel}.
We will instead perform  our calculations in schemes where all soft
contributions are properly factorized into the parton distributions. The
first moment of the gluon coefficient function at NLO is then
{}~\ref\alros{G.~Altarelli and G.~G.~Ross, \PL\vyp{B212}{1988}{391}.}
\eqn\glucoup
{C_g(1,\alpha_s) =-\frac{\as}{4\pi}+O(\alpha_s^2).}
With this choice, the first moment of $g_1$ is given at NLO by
\eqn\gfm{\Gamma_1(Q^2)\equiv \int_0^1 g_1(x,Q^2)dx
=\smallfrac{\langle e^2\rangle}{2}
\left[(1-\smallfrac{\alpha_s}{\pi})
\big(\Delta q_{\rm NS}(1,Q^2) + \Delta \Sigma(1,Q^2)\big)-n_f
\smallfrac{\alpha_s}{2\pi}\Delta g(1,Q^2)\right],}
and the first moment of the scale dependent
eigenvector of the singlet Altarelli-Parisi equations~\aps\
is
\eqn\axfmom{a_0(Q^2)=\Delta \Sigma(1,Q^2)-n_f{\alpha_s\over 2\pi}
\Delta g(1,Q^2).}
The corresponding eigenvalue of the anomalous dimension
matrix coincides with the anomalous dimension of the singlet
axial current, so  $a_0$ is identified with the singlet axial charge
\eqn\achar{\langle p,s|j^\mu_5|p,s\rangle
=M s^\mu a_0(Q^2),}
where $p$, $M$ and $s$ are the momentum, mass and spin of the nucleon.
The other eigenvector of perturbative evolution is
the first moment of the polarized singlet quark distribution,
$\Delta\Sigma(1,Q^2)$, which  is then independent of $Q^2$, and may  be
identified with the conserved singlet quark helicity~\nref\sfpol{S.~Forte,
\NP\vyp{B331}{1990}{1}.}\nref\shve{G.M.~Shore and G.~Veneziano,
\PL\vyp{B244}{1990}{75}; \NP\vyp{B381}{1992}{23}.}\refs{\alros-\shve}.

In fact the eigenvectors remain the same to all orders in
perturbation theory, because of the Adler-Bardeen theorem~\ref\adbar{S.~Adler
and W.~Bardeen, \PR\vyp{182}{1969}{1517}.}, which states
that the NLO mixing of the divergence of the singlet axial current
with a gluonic operator (the anomaly),
which is responsible for its scale dependence, does not receive higher order
corrections. Thus if we require that $\Delta \Sigma(1,Q^2)$
is scale independent then the axial charge is given by eq.~\axfmom\
to all orders. This means that
the first moments of the singlet quark and gluon coefficient functions
are not actually independent: to all orders in perturbation theory
\gfm\ becomes simply
\eqn\gfmao{\Gamma_1(Q^2)=
\smallfrac{\langle e^2\rangle}{2}
\left[C_{\rm NS} (1,\alpha_s)
\Delta q_{\rm NS}(1,Q^2) +C_{\rm S} (1,\alpha_s)a_0(Q^2)\right],}
and thus
\eqn\sgab{C_g (1,\alpha_s)=-\frac{\as}{4\pi}C_{\rm S} (1,\alpha_s).}

Given anomalous dimensions and coefficient functions
in a particular factorization scheme any other factorization scheme can be
constructed~\ref\FP{W.~Furmanski and R.~Petronzio,
\ZP\vyp{C11}{1982}{293}.}  by introducing a scheme change specified by
a function $z_{\rm NS}(N,\alpha_s)=1+{\alpha_s\over 2\pi} z^{(1)}_{\rm NS}(N)
+O(\alpha_s^2)$ and a matrix $z_{\rm S}(N,\alpha_s)=\1+
{\alpha_s\over 2\pi} z_{\rm S}^{(1)}(N)+O(\alpha_s^2)$.
The NLO anomalous dimensions and
coefficient functions then change according to
\eqnn\schad\eqnn\schcf
$$\eqalignno{\gamma_{\rm NS}^{(1)}(N)&\to\gamma_{\rm NS}^{(1)}(N)-
\smallfrac{\beta_0}{2} z_{\rm NS}^{(1)}(N),\cr
\gamma_{\rm S}^{(1)}(N)&\to
\gamma_{\rm S}^{(1)}(N)+[ z_{\rm S}^{(1)}(N),\gamma_{\rm S}^{(0)}(N)]-
\smallfrac{\beta_0}{2} z_{\rm S}^{(1)}(N),&\schad\cr
C_{\rm NS}^{(1)}(N)&\to C_{\rm NS}^{(1)}(N)-z_{\rm NS}^{(1)}(N),\cr
C_{\rm S}^{(1)}(N)&\to C_{\rm S}^{(1)}(N)-z_{qq}^{(1)}(N),&\schcf\cr
C_g^{(1)}(N)&\to C_g^{(1)}(N)-z_{qg}^{(1)}(N),\cr}$$
where $\beta_0=11-\smallfrac{2}{3}n_f$ is the one loop coefficient of the
QCD beta function.

For simplicity we will only discuss scheme changes where $z_{\rm NS}=z_{qq}$,
i.e. such that the relative normalization of the singlet and
nonsinglet quark distributions is unaffected.
The conservation of the nonsinglet axial current then fixes the
first moment $z_{\rm NS}^{(1)}=z_{qq}^{(1)}=0$ whenever the original
scheme respects chiral symmetry.
We then wish to consider specifically factorization schemes in which
the first moments of the coefficient functions
satisfy eq.~\glucoup: starting from the \MS\ scheme,
eq.~\schcf\ then fixes $z_{qg}^{(1)}(1)=1$.\foot{A
scheme change of this kind was constructed
in ref.~\ref\ziner{ E.B.~Zijlstra and
W.L.~van~Neerven, \NP\vyp{B417}{1994}{61}.}; the form of the matrix $z$ given
there appears however to be incorrect. Also, note that the partial result
for the NLO splitting functions given there is incorrect, as explained
in ref.~\merner .}
In order to completely specify the first moment of
$z$ we use the Adler-Bardeen condition that the
two-loop eigenvector of perturbative evolution as given by \axfmom\
be identified to all perturbative orders with the matrix element of
the axial current. Knowledge of the NNLO anomalous dimension of the axial
current~\ref\larin{S.~A.~Larin, {\it Phys. Lett.} {\bf B334} (1994) 192.}
then fixes the first moments of the remaining two entries of the matrix
$z$.

We will consider several schemes which differ in the way the remaining
moments of the coefficient functions and anomalous
dimensions are constructed. In the first scheme, we
simply take $z(x)$, the inverse Mellin transform of $z(N)$,
to be independent of $x$.  This scheme is thus the minimal modification
of the \MS\ scheme such that the first moments of parton distributions satisfy
the anomaly constraint eq.~\axfmom; in particular,
the large and small $x$ behaviour of the coefficient functions and anomalous
dimensions are then the same as in \MS.
We will refer to this as the Adler-Bardeen (AB) scheme. The matrix
which transforms~from \MS\ to the AB scheme is
\eqn\zab
{z_{\rm S}^{(1)}(N)\big\vert_{\rm AB}={1\over N}
\pmatrix{&0& 2 n_f T_F \cr &0&0\cr},}
where, for SU(3) color, $C_F={4\over3}$, $C_A=3$ and $T_F={1\over2}$.
Notice that the two lower entries in the \MS\ NLO anomalous dimension matrix
\merner\ turn out to be already
consistent with the Adler--Bardeen  condition above,
and NNLO anomalous dimensions~\larin; the corresponding entries of the
scheme change matrix eq.~\zab\ therefore vanish.

Transformations such as \zab\ which take us from \MS\ to a scheme where the
gluon contributes to the first moment of $g_1$ correspond to removing
soft contributions from the coefficient functions~\refs{\alam-\vogel}.
Rather than doing this by hand, as in the AB scheme above, the
subtraction may be performed by computing the coefficient functions
in the presence  of an explicit infrared regulator, which automatically
enforces eq.~\glucoup~\vogel. The entries
$z_{qq}$ and $z_{qg}$ of the $z$ matrix are then fixed using
eq.~\schcf\ and the \MS\ coefficient functions~\merner
\eqn\MScf{\eqalign{{C_q^{\rm S}}^{(1)}(N)\big\vert_{\overline{\rm MS}}&=
C_F\Big[S_1(N)\left(\smallfrac{3}{2}-\smallfrac{1}{N( N+1)}+S_1(N)\right)
{}~- S_2(N)-\smallfrac{9N^3+6N^2-3 N-2}{2 N^2(N+1)}\Big],\cr
{C_g^{\rm S}}^{(1)}(N)\big\vert_{\overline{\rm MS}}&=-T_F
\smallfrac{N-1}{N(N+1)}\left(S_1(N)+
\smallfrac{N-1}{N}\right),\cr}}
where $S_j(N)=\sum_{k=1}^N {1\over k^j}$.
The two lower entries of the transformation matrix
can then be taken to be zero as in the AB scheme.

One possibility is to renormalize while keeping the incoming particle
off-shell (OS scheme, henceforth);
the coefficient functions are then given by
\eqn\OScf{\eqalign{{C_q^{\rm S}}^{(1)}(N)\big\vert_{\rm OS}&=
C_F\left(\smallfrac{3}{2}S_1(N)-4S_2(N)
{}~-\smallfrac{2N^4-N^3-5N-4}{2 N^2 (N+1)^2}\right),\cr
{C_g^{\rm S}}^{(1)}(N)\big\vert_{\rm OS}&
=-2T_F\smallfrac{N^3-N^2+N+1}{N^2(N+1)^2}.\cr}}
An alternative option is to endow the quarks with a finite mass
(Altarelli-Ross, or AR scheme);
the quark~\ref\strat{M.~Stratmann, A.~Weber and W.~Vogelsang, Dortmund
preprint DO-TH 95/15, {\tt hep-ph/9509236}}\foot{Because the quark
mass breaks chiral symmetry, it is now necessary to perform an extra
subtraction in order to ensure that the nonsinglet axial currents
are conserved.} and
gluon~\refs{\alros,\vogel} coefficient functions are then
\eqn\OScf{\eqalign{{C_q^{\rm S}}^{(1)}(N)\big\vert_{\rm AR}&=
C_F\Big[\left(
\smallfrac{7}{2}+\smallfrac{1}{N(N+1)}-S_1(N)\right)S_1(N)
{}~-3S_2(N)-\smallfrac{5N^4+7N^3+5N^2-3N-2}{2 N^2 (N+1)^2}\Big],\cr
{C_g^{\rm S}}^{(1)}(N)\big\vert_{\rm AR}
&=-T_F\smallfrac{N^2+1+N(N-1)S_1(N)}{N^2(N+1)}.\cr}}
Notice that in all three of these schemes the first moments of the coefficient
functions are given by eqs.~\qucoup\ and \glucoup, and thus the NLO
relation between the first moment of $g_1$ and the singlet axial
charge $a_0$ implicit in eqs.~\gfm\ and \axfmom\
is automatically satisfied.

The main effect of the NLO corrections to perturbative evolution is the
coupling of the gluon to $g_1$, and in particular its contribution to the
first moment $\Gamma_1$ eq.~\gfm, which does not decouple as
$Q^2\to\infty$~\alros. However, NLO corrections may also
substantially affect the small-$x$ behaviour of parton distributions and
coefficient functions. Indeed, unlike the unpolarized case,
the NLO contributions to the polarized splitting functions
and coefficient functions display a stronger singularity as $x\to 0$ than
their LO counterparts; accordingly their Mellin transforms display a
stronger singularity as $N\to 0$. More specifically the singularities
in the NLO \MS\ anomalous dimensions~\merner\ and coefficient
functions take the form\eqnn\smallnad\eqnn\smallncf
$$\eqalignno{\gamma_{\rm NS}^{(1)}(N)&={1\over N^3}\left(2 C_A C_F - 3
C_F^2\right)+O\left({1\over N^2}\right),&\smallnad\cr
\gamma_{\rm S}^{(1)}(N)&={1\over N^3}\pmatrix{&- 4 C_F T_F n_f
+ 2 C_A C_F- 3 C^2_F& -2 C_A T_F  -  C_F T_F \cr
& 4 C_A C_F + 2 C_F^2 & 8 C_A^2-4 C_F T_F n_f\cr}+O\left({1\over N^2}\right),
\cr
C_{\rm NS}^{(1)}&=C_q^{(1)}= C_F{1\over N^2} +O\left({1\over N}\right),
\qquad C_g^{(1)}=- T_F{1\over N^2}+O \left({1\over N}\right),&\smallncf
\cr}$$
whereas the LO anomalous dimensions have a simple pole in $N$ (details
of which may be found in ref.~\BFR).\foot{The presence of double
logarithms in the NLO polarized splitting functions and coefficient
functions strongly suggests that the systematic cancellation of
collinear singularities which characterizes the
small-$x$ behaviour of unpolarized splitting functions does not
occur in the polarized case. In the nonsinglet channel a summation
of these double logarithmic singularities to all orders in
$\alpha_s$ has been attempted in
ref.~\ref\bar{J.~Bartels, B.~I.~Ermolaev and M.~G.~Ryshkin, DESY
preprint  95-124.}.}

The NLO corrections could therefore have a significant impact in
the small $x$ region, and in particular require a summation of
logarithmic effects in $1\over x$, which could be done in analogy to the
unpolarized case~\ref\summ{R.~D.~Ball and S.~Forte, {\it Phys. Lett.}
{\bf B351} (1995) 313.} if the coefficient of the most singular
contributions to the polarized splitting functions as $x\to0$ were known
to all orders in $\alpha_s$. In order to assess at which values of
$x$ and $Q^2$ these effects might begin to be relevant (and in particular
whether they already amount to a sizable correction in the
presently measured region) $g_1$ may be
determined by solving the NLO evolution
equations with the approximate small-$N$ form eq.~\smallnad\
of the anomalous dimensions:
\eqn\sxNLO{\eqalign{\Delta q_{\rm NS}(N,Q^2)=&\Delta q_{\rm NS}(N,Q^2_0)
\left({\alpha_s(Q^2_0)\over \alpha_s(Q^2)}\right)^{{2\gamma_{\rm NS}^{(0)}(N)
\over \beta_0}}\left[1+{\epsilon_{\rm NS}\over N^3}
\left({\as(Q^2_0)}-{\as(Q^2)}\right)\right],\cr
v_\pm(N,Q^2)=&v_\pm(N,Q^2_0)\left({\alpha_s(Q^2_0)\over \alpha_s(Q^2)}
\right)^{{2\lambda_\pm(N)\over \beta_0}}
\left[1+{\epsilon_\pm\over N^3}
\left({\as(Q^2_0)}-{\as(Q^2)}\right)\right],\cr}}
where $v_\pm$ and $\lambda_\pm$ are the eigenvectors and eigenvalues
of the LO singlet anomalous dimension matrix $\gamma^{(0)}$ (see ref.~\BFR),
$\alpha_s(Q^2)$ is computed at NLO, $Q_0$ is the
starting scale, and the coefficients
$\epsilon$ are explicitly given by
\eqn\delco{\epsilon_{\rm NS}= \smallfrac{8}{3\pi\beta_0},\qquad
\epsilon_\pm =\smallfrac{112}{3\pi\beta_0}\bigg[
(1-\smallfrac{n_f}{14})\pm
\smallfrac{13}{14}(1-\smallfrac{11n_f}{104})
\Big/\sqrt{1-\smallfrac{3n_f}{32}}\bigg].}
The NLO corrections do not mix the LO small $x$ eigenvectors, because
mixing terms are $O\left({1\over N^2}\right)$.
The eigenvectors of perturbative
evolution at NLO are thus the same as at LO:
at small $x$ and large $Q^2$ $\Delta g$ and $\Delta \Sigma$
have opposite sign~\BFR, and in particular (for any plausible parton
distributions)
$\Delta \Sigma<0$ and $\Delta g>0$.
It is interesting to observe that this result is scheme independent,
because the leading $1\over N^3$ singularities
in the NLO corrections only receive contributions from the diagonal
projections
of the NLO anomalous dimension matrix onto the LO eigenvectors, which are
themselves scheme independent.

The leading NLO small-$x$ behaviour can now be
found by inverse Mellin transform of eq.~\sxNLO\ in the saddle point
approximation:
\eqn\sxxNLO{\eqalign{\Delta q_{\rm NS}(x,Q^2)=&{\cal N}_{\rm NS}
\sigma^{-1/2}e^{2\gamma_{\rm NS}\sigma}
\Big[1+\epsilon_{\rm NS}
\big(\smallfrac{\rho}{\gamma_{\rm NS}}\big)^3
\left({\as(Q^2_0)}-{\as(Q^2)}\right)\Big],\cr
v_\pm(x,Q^2)=&{\cal N}_\pm
\sigma^{-1/2}e^{2\gamma_{\pm}\sigma}
\Big[1+\epsilon_{\pm}
\big(\smallfrac{\rho}{\gamma_\pm}\big)^3
\left({\as(Q^2_0)}-{\as(Q^2)}\right)\Big],\cr}}
where $\cal N$ are normalization constants,
$\sigma\equiv\sqrt{\xi\zeta}$, $\rho\equiv\sqrt{\xi/\zeta}$,
$\xi\equiv\ln\smallfrac{x_0}{x}$,
$\zeta\equiv\ln\smallfrac{\alpha_s(Q_0^2)}{\alpha_s(Q^2)}$,
$x_0$ is a reference value of
$x$ such that the approximate small-$x$ form of the anomalous dimensions
is applicable for $x\lsim x_0$ and $Q^2\gsim Q^2_0$, and
$\gamma_{\rm NS}^2= {2C_F\over \beta_0}$ while $\gamma_\pm$ are as given
in ref.~\BFR.\foot{Eq.~\sxxNLO\ only gives the NLO generalization
of the asymptotic LO behaviour $\sigma^{-1/2}e^{2\gamma\sigma}$ when
the boundary conditions are soft, as discussed in ref.~\BFR.
If the boundary condition is hard, for example
$e^{\xi\lambda}$, then for
$\rho\gsim\gamma/\lambda$  the LO
behaviour reproduces the boundary condition~\BFR, and the NLO
correction to it is given by
$\left[1+\epsilon_i\lambda^{-3}
\left({\as(Q^2_0)}-{\as(Q^2)}\right)\right]$, where
$\epsilon_i=\epsilon_{\rm NS},\epsilon_\pm$ in
in the nonsinglet and singlet cases respectively.
In this case the NLO correction is thus $x$-independent.}
The terms in square brackets in eq.~\sxxNLO\ give the NLO correction
to the LO asymptotic small $x$
behaviour\nref\polscal{M.~A.~Ahmed and G.~G.~Ross,
{\it Phys. Lett.} {\bf B56} (1975)
385\semi M.~B.~Einhorn and J.~Soffer, {\it Nucl. Phys.} {\bf B74} (1986) 714
\semi A.~Berera, {\it Phys. Lett.} {\bf B293} (1992) 445.}
\refs{\BFR,\polscal}. Because
all coefficients $\epsilon$ eq.~\delco\ are positive the NLO corrections lead
to a further increase proportional to $\xi^{3/2}$ of the parton
distributions at small $x$. The coefficient of this increase is however rather
small, for instance with $n_f=4$ one gets
${\epsilon_{\rm NS}/\gamma_{\rm NS}^3}\approx
{\epsilon_+/\gamma_+^3}\approx\half$,
so that the correction is small
in the presently accessible small $x$ region.
These conclusions however only apply
to the region where the no summation of logs of $1\over x$ is necessary
so that the NLO in $\alpha_s$ may be treated as a subleading correction, and
could be substantially altered at smaller values of $x$.

The leading small $x$ behaviour of $g_1$ can be found at NLO using
the small $x$ NLO solution eq.~\sxNLO\ and coefficient functions
eq.~\smallncf\ in the Mellin transform of the
expression of $g_1$ eq.~\gone. Because the NLO
coefficient functions~\smallncf\ only have a $1\over N^2$ singularity at NLO
they actually do not contribute to the leading small-$x$ behaviour of $g_1$,
which is thus found by simply taking the appropriate linear combination of
the small-$x$ parton distributions eq.~\sxxNLO.
The NLO correction to the small-$x$ behaviour of the coefficient functions may
nevertheless have a significant impact, especially at low scales (i.e.,
when $Q$ is close to the starting scale $Q_0$) where evolution effects are
negligible. Indeed, the $1\over N^2$ singularity corresponds to a
$\log{1\over x}$ rise of the coefficient function, and would therefore lead
to a rise of both singlet and nonsinglet contributions to the
structure function $g_1$ even if
the parton distributions themselves did not rise. The coefficient
of this rise is however not scheme
independent: for instance, the coefficient of the leading singularity in the
gluon coefficient function is the same in the \MS, AB and AR schemes, but
is twice as large in the OS scheme.

Having established that a NLO treatment is adequate in the region
of current experimental data, we can proceed to a determination
of the physical observables related to $g_1$.
Even though our purpose here is not to establish a parametrization
of polarized parton distributions, we have to construct such a
parametrization since only LO  parametrizations are
currently available\foot{A comprehensive review of the present status
of polarized parton parametrizations is given in
ref.~\ref\ppar{T.~Gehrmann and
W.~J.~Stirling, Durham preprint DTP/95/78, {\tt hep-ph/9510243}.}. A
NLO parametrization has been presented in
ref.~\ref\glure{M.~Gl\"uck et al., Dortmund preprint DO-TH 95-13 {\tt
hep-ph/9508347}.}, but in
the unsubtracted \MS\ scheme, which, as previously discussed, is
not properly factorized.}. We parametrize the initial parton
distributions according to
\eqn\parm{\Delta f(x, Q_0^2)=N\left(\alpha_f,
\beta_f,a_f\right)
\eta_fx^{\alpha_f}
(1-x)^{\beta_f}(1+a_fx),}
where  $N(\alpha,\beta,a)$ is fixed by the normalization condition
$N(\alpha,\beta,a)\int_0^1\!dx\, x^{\alpha}
(1-x)^{\beta}(1+ax)=1$, and $\Delta f$ denotes $\Delta \Sigma$,
$\Delta q_{\rm NS}$ or
$\Delta g$.
In our previous analysis~\BFR, which used only proton data, the
respective three sets of
parameters  could not be independently determined,
while the small $x$ behaviour
had to be fixed and then varied in a plausible range.
Now, by including the deuteron data~\deuex, we can disentangle
the nonsinglet and singlet quark and gluon
contributions to $g_1$, because
$g_1^p$ is dominated by the isotriplet term, which contributes
about 90\% of its first moment, while $g_1^d$ is isosinglet. We can
thus determine independently almost all of the parameters of
the singlet and nonsinglet parton distributions (including the
parameters $\alpha_f$, which determine their small-$x$ behaviour).

We determine $g_1$ at all $x$ and $Q^2$ by solution of the
NLO evolution equations with boundary conditions of the form~\parm\
at $Q_0^2=1$~GeV$^2$.
The various parameters in eq.~\parm\ are then found
by fitting $g_1(x,Q^2)$ to the recent precision experimental
determination of $g_1$ for proton and deuteron~\refs{\proex,\deuex},
which are given along a curve $Q^2=Q^2(x)$
for each experiment.  As in \BFR\ $g_1$ is extracted
{}~from the measured asymmetry $A_1$ using the leading-twist
expression\foot{Kinematic higher twist corrections which
are sometimes included are neglected here for consistency
since no systematic treatment of these corrections is available.}
$g_1(x,Q^2)= A_1(x,Q^2)F_1(x,Q^2)$; $F_1$ is in turn computed
{}~from the SLAC determination~\ref\Rslac{L.~W.~Whitlow et al.,
\PL\vyp{B250}{1990}{193}.} of the ratio $R(x,Q^2)$ of the
longitudinal to transverse virtual photoabsorption cross section,
and the most recent NMC determination~\ref\NMC{NMC Collaboration, M.~Arneodo
et al. preprint CERN-PPE/95-138, {\tt hep-ph/9509406}.} of
$F_2(x,Q^2)$. The deuteron structure function is defined as the average
of the proton and neutron structure functions and is obtained from
the deuteron asymmetry after applying a  correction to account for
$d$-wave
admixture~\ref\dwav{L.~L.~Frankfurt and M.~Strikman,
{\it Nucl. Phys.} {\bf A405} (1983) 557.}:
\eqn\deustf
{g_1^d(x,Q^2)\equiv\half\big(g_1^p(x,Q^2)+g_1^n(x,Q^2)\big)
=A_1^d(x,Q^2) F_1^d(x,Q^2)\big/(1-1.5\omega_D)}
where $\omega_D=0.05$.

The normalization of the nonsinglet quark distribution
at $Q_0$ (which lies below the charm threshold) is
fixed by assuming SU(3) symmetry of the matrix
elements of the axial current,
determined~\ref\CR{F.~E.~Close and R.~G.~Roberts,
\PL\vyp{B336}{1994}{257}.} from hyperon $\beta$
decays:\eqnn\normns\eqnn\nsval
$$\eqalignno{\eta_{\rm NS}&=\int_0^1 \Delta q_{\rm NS}(x,Q^2)dx=\pm
\smallfrac{3}{4}g_A+\smallfrac{1}{4}a_8,
&\normns\cr
g_A&=1.2573\pm 0.0028;\qquad a_8=0.579\pm 0.025,&\nsval\cr}$$
where the plus (minus) sign refers to a proton (neutron) target.
While the impact of possible SU(3) violation on our results
will be discussed below, we will defer to a subsequent publication the
possibility of using the data themselves to
test SU(3) or SU(2) (and thus the Bjorken sum rule).
At higher scales, new nonsinglet contributions arise as
the various heavy quark thresholds are passed, so that
$\Delta q_{\rm NS}(1,Q^2)$
is not scale independent across thresholds.
The corresponding
heavy quark distributions are
generated dynamically, assuming that they
vanish on threshold,
and imposing~\alam\ continuity of $a_0(Q^2)$ eq.~\axfmom.

It turns out that the data are good enough to determine all the remaining
parameters directly with a single exception, namely, the parameters $a_q$ and
$a_g$ which control the shape of the singlet distributions $\Delta
\Sigma$ and $\Delta g$ at intermediate $x$, and which are difficult
to pin down individually since
the evolution mixes $\Delta \Sigma$ and $\Delta g$. We will thus
take $a_q=a_g$, and determine the remaining ten parameters from a fit
to the data. The respective best fit values  are listed, in
table~1 for the AB, AR and OS schemes. The errors
given there are statistical errors from the fit, computed by  taking
into account  correlations.
The corresponding determination of
$g_1(x)$ in the AB scheme is shown in fig.~1 at the starting scale
as well as at
the scale $Q(x)$ of the various data sets and at $Q^2=10$~GeV$^2$.

The main features of this determination of polarized parton distributions are
the following. The evolution of $g_1$ is rather similar to that discussed in
ref.~\BFR, thus demonstrating that the main NLO effect is indeed the
direct gluon
contribution to the first moment of $g_1$ which was already included there; the
main difference between the present  determination of polarized parton
distributions and that of ref.~\BFR\ is the more detailed nature of
the fit, due to the inclusion of the deuteron data.

Specifically, the large $x$ behaviour, controlled by the exponent
$\beta$, is in
fair agreement with the expectations based on QCD counting rules~\ref\brod{See
S.~J.~Brodsky, M.~Burkardt and I.~Schmidt, {\it Nucl.Phys.} {\bf B441} (1995)
197.}, which predict $\beta_q=\beta_{\rm NS}\simeq3$ and
$\beta_g\simeq 4$; of
course these parameters are strongly scheme dependent (as explicitly displayed
in table 1). The value of $\beta_g$ cannot be determined very accurately since
large-$x$ data are at relatively large $Q^2$, where the direct coupling of the
gluon to $g_1$ is suppressed by the small value of $\alpha_s$; in the OS scheme
the results are  largely independent of this coefficient which
thus cannot be  fitted at all.

The small $x$ behaviour turns out to be at least qualitatively constrained
by the data. The initial singlet parton distributions are found to
display soft valence-like behaviour, i.e. not to grow at small $x$,
as was predicted by Regge theory~\ref\heim{R.~L.~Heimann,
\NP\vyp{B64}{1973}{429}.}. However the nonsinglet quark distribution
is found to grow approximately as $\Delta q \neath \sim {x\to 0}
{1\over\sqrt{x}}$.\foot{This seems to be qualitatively consistent
with the summation of double logarithms presented in \nref\emr{B.I.~Ermolaev,
S.I.~Manayenkov and M.G.~Ryskin, DESY preprint 95-017.}
refs.\refs{\bar,\emr}, since this suggests that both
the polarized and unpolarized nonsinglet quark distributions have the
same leading small $x$ behaviour, at least in perturbation theory.} We have
explicitly verified that these behaviours
indeed correspond to an overall global minimum, and that the quality
of the fit deteriorates uniformly as the values of $\alpha_f$ deviate from
those given in table 1, by repeating the fit with fixed values of
the three exponents $\alpha_f$ chosen in the range $-0.9\le \alpha_f \le +0.5$.
The singular behaviour of the nonsinglet quark is the main difference between
these results and those discussed in ref.~\BFR, where, since it was
impossible to determine it from the data, it was assumed to be
valence-like.

Even though these results are suggestive, they should be taken
with some care: firstly, the values of $\alpha_f$ are strongly
scheme-dependent, and partly just reflect the shape of the
coefficient functions at small $x$ and small $Q^2$ in the various schemes:
for instance, the fact that the
coefficient of the leading singularity in $C_g^{(1)}$ is larger
in the OS scheme explains why the value of $\alpha_g$ in this scheme is
larger. Furthermore, these are not necessarily the asymptotic
behaviours of the various parton distributions as $x\to 0$: in particular,
the large value and uncertainty in $a_{\rm NS}$ indicate that the
asymptotic behaviour of $g_1^p$
is still setting in at the smallest experimental $x$
values; likewise, $g_1^d$ at small $Q^2$ and small $x$ grows large and positive
with the present parametrization, but this happens outside the data range
(the growth is barely seen starting on the $Q^2=1$~GeV$^2$ curve of Fig.~1b).
In this respect, present-day polarized data are to be compared to the NMC
unpolarized data~\NMC: a kinematic
coverage comparable to that available at HERA for
determining~\ref\das{R.~D.~Ball and S.~Forte, \PL\vyp{B335}{1994}{77}.} the
small-$x$ behaviour $F_2$ would be
required to determine precisely the small $x$ behaviour
of polarized parton distributions. This is reflected by the large
uncertainties in the parameters $a_f$ which govern the transition to the
small $x$ region.

Finally, the data turn out to allow a good determination of the size
of the quark and gluon distributions. The size of the gluon distribution drives
perturbative evolution; the evolution already observed due to the fact that
the SMC and E143 data are taken at different values of $Q^2$ for equal
$x$ (displayed in fig.~1) turns out to be sufficient to require
$\eta_g$ (which gives the first moment of $\Delta g$ at the initial scale
$Q=Q_0$) to be large and positive.
The relation eq.~\axfmom\ between the first moments of the quark
and gluon distributions and the first moment of $g_1$ then
inevitably leads to a rather large quark singlet.
This is a remarkable result: the scale-independent first moment of $\Delta
\Sigma$, given by $\eta_q$, is found empirically to be equal to
$a_8$ eq.~\nsval\ within errors. The large violation of the Zweig
rule in the first moment of $g_1$ appears then to be almost entirely
due to a large perturbative gluon contribution to $g_1$,
as was conjectured in ref.~\alros.

A large gluon distribution implies substantial evolution effects, and
thus a substantial correction due to the determination of the moments
of $g_1$ from the experimental data~\BFR.
Indeed, we can now determine the first moment eq.~\gfm, as well as
the singlet axial
charge of the nucleon defined according to eq.~\axfmom\ from the  best-fit
polarized parton distributions. The results in the AB scheme, which we
shall take as a baseline, are displayed in the first row of table~2. Only
$\Gamma_1^p$ is shown; the deuteron is obtained from it by
subtracting
the isotriplet contribution $\Gamma_1^{I=1}=\smallfrac{1}{12}
C_{\rm NS}(1,\alpha_s) g_A$ since we have assumed isospin to be exact.
The value of
$\Gamma_1$ is found to be lower than the value extracted~\refs{\proex,\deuex}
neglecting the evolution of the asymmetry, and consequently the value
of the axial charge $a_0$ we obtain is about a half of the value
quoted by the experimental collaborations. It
is thus important to establish whether this result is a robust consequence of
the inclusion of NLO evolution effects, or whether it is a by-product of the
procedure used so far. Furthermore, if the result is confirmed, it inevitably
implies an increase in the theoretical uncertainty in the determination of the
axial charge, which must be more accurately estimated.

First, we test the sensitivity to the specific
functional form eq.~\parm\ of the parton distributions. To this purpose,
we change the parameterization \parm\ by replacing the last factor
with $(1+b_f\sqrt{x}+a_fx)$, and then repeating the fit with
various choices for the new parameters $b_f$.
In particular, we fix $b_q=b_g$ and then we vary $-5\le b_q,\;
b_{\rm NS} \le 5$. No significant variation in the results
or the quality of the global fit
is found, indicating that the fit is stable. The results
for extreme values of $b_{\rm NS}$ while $b_q=b_g=0$ are shown
in table~2; variations of $b_q$, $b_g$ produce even smaller effects.

Other sources of uncertainty related to the fitting procedure are due to
the fact that the structure function is not measured over the full
range of values of $x$, and is only sampled at
a finite number of points.
The former uncertainty comes from the extrapolations to small and
large $x$, and is thus already included through the propagated errors
in the six parameters $\alpha_f$ and $\beta_f$. It is interesting to
note that if  we were to assume that the initial nonsinglet quark
distribution was flat at small $x$, i.e. $\alpha_{\rm NS}=0$,
$\Gamma_1$ would scarcely be altered, even though $\eta_q$ and $\eta_g$ both
fall substantially (see table~2). The latter effect, however  is due more to
the deterioration in the
quality of the fit at intermediate $x$ than to a large change in the
small $x$ contribution.  The other
uncertainty is related to the fact that we determine the first moments
of $g_1$ and the parton distributions by integrating the
respective best-fit forms. This is to be contrasted with the procedure
followed by experimental collaborations, which instead determine
the first moment of $g_1$ by summing the data over the experimental
bins; evolution effects could then be included as
corrections to each bin separately. Such a procedure is however problematic
because evolution effects are actually rather large in many of the bins,
and furthermore the value of $g_1(x_0,Q_0^2)$
is determined (nonlinearly) by the measurements at all $x>x_0$
and $Q^2<Q_0^2$ so it is not possible to disentangle truly independent
corrections to individual data points. Thus a simple sum over
bins ignores the constraints imposed by perturbative evolution on the
shape of $g_1$ in the $(x,Q^2)$ plane and gives undue weight to
particular data points. The integration of fitted distributions
has nevertheless an inevitable sampling uncertainty. To estimate this
we have computed the difference between the
experimental value of $g_1$ in each bin and our best-fit $g_1(x_b,Q_b^2)$,
where $(x_b,Q^2_b)$ are the central values of $x$ and $Q^2$
for each $x$ bin, and have then computed the first moment $\Delta$
of this difference over the measured region by multiplying it by the bin
width and summing over bins. We find $\Delta^p({\rm SMC})=-0.0032$,
$\Delta^p({\rm E143})=0.0007$, $\Delta^d({\rm SMC})=-0.0045$,
$\Delta^d({\rm E143})=0.0010$
for the four experiments, showing that this correction is indeed small and
does not lead to a systematic bias. Combining these results we
obtain an estimate for the overall  statistical uncertainty due to the
choice of the fitting procedure, to be added to that of the fit itself
(see table~3).

We now turn to the various sources of systematic theoretical uncertainty.
All the corresponding results, along with the $\chi^2$ of the various
fits, are summarized in table 2. First, we consider
the possibility of SU(2) or SU(3) violation. The former will in general
induce current mixing effects: for instance, in the presence
of isospin violation, the proton and deuteron matrix element of the isosinglet
current may  be unequal to each other.
However, due to the expected smallness
of isospin violation, this effect is negligible compared to the error on
the values of $g_A$ and $a_8$ induced by SU(2) and SU(3) violation
in the quark distributions,
which are then the dominant source of uncertainty. We estimate these by
assuming a 2\% uncertainty on $g_A$
(of the same order of the accuracy to which isospin symmetry of unpolarized
quark distributions may be established~\ref\iso{S.~Forte,
\PR\vyp{D47}{1993}{1842}.}) and a 30\% uncertainty on
$a_8$~\ref\suthree{B.~Ehrnsperger and A.~Sch\"afer,
\PL\vyp{B348}{1995}{619}\semi J.~Lichtenstadt and H.~J.~Lipkin,
\PL\vyp{B353}{1995}{119}.}. Notice that when these parameters are varied
the size of the nonsinglet first moment of $g_1$ varies; however,
table~2 shows that the best-fit value of the singlet first moment and
thus of $a_0$ adjusts itself and remains surprisingly stable.

The uncertainty related to the value of $\as$, which
is not negligible in a NLO computation, is simply estimated by
repeating the fit when  $\as$ is varied
in the range~\ref\webas{B.~Webber, summary talk at the International
Conference on High Energy Physics (ICHEP), Glasgow, Scotland, 1994,
{\tt hep-ph/9410268}.} $\alpha_s(M_z)=0.117\pm0.005$.
The position of the quark thresholds is varied from $0.75 m_q$ to $2.5
m_q$ (with $m_c=1.5$~GeV, $m_b= 5$~GeV).

An important source of theoretical uncertainty is due to the lack of knowledge
of higher order corrections, as reflected by the dependence of the
results on factorization scale $M^2$ and renormalization scale
$\mu^2$. We estimate this by taking $M^2=k_1 Q^2$ and
$\mu^2=k_2 Q^2$, and varying $0.5\le k_1, k_2\le 2$.
When the scales are varied the physical parameters
turn out to have a stationary point within this range; the
associated error is thus asymmetric, but rather large,
consistent with the fact that evolution effects are important.
The fluctuations found when
varying the factorization scale are consistent with the spread of the
results displayed in table~1 in various factorization schemes. We have
also checked that including the known two and three loop corrections
to the first moments of the coefficient functions produces similar variations.

The errors corresponding to all these sources, given by the maximal
variation of the results as the parameters are varied in the
respective ranges, are summarized in table~3. We have not included an
error due to higher twist terms since we know of no reliable way of
estimating it: experience with unpolarized data suggests that it is
probably rather smaller than the error from higher order corrections.

In conclusion, the analysis of various sources of theoretical uncertainty
confirms the stability of our determination of the first moment of
$g_1$ and the axial charge of the nucleon. Adding in quadrature the
various sources of error, summarized in table 3, we get finally
\eqn\citterio
{\eqalign{
\Gamma_1^p(3~{\rm GeV}^2)&=0.118\pm 0.013\>\hbox{(exp.)}
\epm{0.009}{0.006}\>\hbox{(th.)},\cr
\Gamma_1^d(3~{\rm GeV}^2)&=0.024\pm 0.013\>\hbox{(exp.)}
\epm{0.011}{0.005}\>\hbox{(th.)},\cr
\Gamma_1^p(10~{\rm GeV}^2)&=0.122\pm 0.013\>\hbox{(exp.)}
\epm{0.011}{0.005}\>\hbox{(th.)},\cr
\Gamma_1^d(10~{\rm GeV}^2)&=0.025\pm 0.013\>\hbox{(exp.)}
\epm{0.012}{0.004}\>\hbox{(th.)},\cr
a_0(10~{\rm GeV}^2)&=0.14\pm 0.10\>\hbox{(exp.)}
\epm{0.12}{0.05}\>\hbox{(th.)},\cr}}
The deuterium values refer to the structure function eq.~\deustf;
they must be multiplied by $1-1.5\omega_D$ in order to compare
with the results quoted by the experimental collaborations.
Values of $a_0$ at any other scale are obtained using NLO
evolution~\koda
\eqn\azev{a_0(Q^2)= \big[1+\smallfrac{2n_f}{\beta_0}
\smallfrac{\alpha_s}{\pi}+O\left(\alpha_s^2\right)
\big]a_0(\infty),}
which the expression \axfmom\ $a_0$ satisfies by construction.
The dominant source
of theoretical uncertainty is that related to higher order corrections, i.e.
to  renormalization and factorization scale. The experimental uncertainty
includes the various errors related
to the fitting procedure (hence also the experimental systematics),
and is still dominant. This is mainly due to the fact
that the data allow us to only partly constrain the shape of the
polarized gluon distribution, and thus the perturbative
evolution of $g_1$.

It is interesting to compare the value of the axial charge eq.~\citterio,
obtained using eq.~\axfmom, with that obtained using eq.~\gfmao\ with
the coefficient functions expanded to NLO as in eq.~\qucoup:
\eqn\achpr{a_0^\prime(Q^2)\equiv [C_{\rm S}(1,\alpha_s)]\inv
\left(\smallfrac{2}{\langle e^2\rangle}\Gamma_1(Q^2)
{}~-C_{\rm NS}(1,\alpha_s)\Delta q_{\rm NS}(1,Q^2)\right).}If we could work to
all orders $a_0=a_0'$; however, in
a $k$-th order perturbative computation  the two determinations differ by
($k+1$)-th order corrections. Indeed, at NLO
\eqn\comp{a_0^\prime(Q^2)-a_0(Q^2)=
{}~-2 n_f\left({\alpha_s\over 2\pi}\right)^2 \Delta g(1,Q^2)+O(\as^3).}
This difference may be quite large if the gluon
distribution is large: indeed, using eq.~\achpr\
we get $a_0^\prime(10~{\rm GeV}^2)=0.09\pm 0.15\>
\hbox{(stat.)}$, which differs considerably from the central value
eq.~\citterio, though within the
theoretical error. Notice that \comp\ is scale dependent:
while it vanishes asymptotically, at low scales it is large
enough that $a_0^\prime(Q^2)$ actually increases with $Q^2$
below $10$~GeV$^2$  rather than decreasing as the axial charge
should, according to eq.~\azev.

Finally, we can estimate the size of the quark and gluon distributions
in the AB scheme:
\eqn\qgres{\eqalign{\int_0^1\!dx\,\Delta \Sigma(x)
&=0.5\pm0.1,\cr
\int_0^1\!dx\,\Delta g(x,1 \hbox{GeV}^2 )
&=1.5\pm 0.8,\cr}}
where all errors have been added in quadrature.
The gluon distribution is rather large even at the quoted low scale
(the corresponding value at 10~GeV$^2$ is roughly twice as large)
and even though the error is large, it differs significantly
{}~from zero. The value of the (scale independent) singlet quark
distribution is in agreement within errors with the prediction of
the Zweig rule, which would identify it with $a_8$ eq.~\nsval.

In summary, we have given a NLO determination of the main physical observables
related to the polarized structure function $g_1$. Our main result is that
the data already constrain the size of polarized parton
distributions, and in particular require a rather large polarized gluon
distribution. This in turn implies that perturbative evolution
effects are not negligible, and in fact substantially affect the
extraction of the first moment of $g_1$ from experimental data.
Indeed, we find a value of the nonsinglet first moment which
is significantly smaller
than that obtained from a purely LO analysis. More importantly, we show that
the error on the determination of the singlet
axial charge of the nucleon is significantly larger than usually
recognized, due essentially to the unknown effects of higher order perturbative
corrections. The recent precise data on $g_1$ are providing us
with surprisingly accurate information on polarized parton
distributions, but they also show that the theoretical and phenomenological
interpretation of these data is significantly more subtle than
previously expected.

\bigskip
{\bf Acknowledgement:} We thank R.~Mertig and W.~van~Neerven for providing
details on their computations, P.~Bosted, R.~Erbacher, B.~Frois and V.~Hughes
for information about the experimental data, and  G.~Altarelli,
R.K.~Ellis and G.~Mallot for discussions.
\vfill\eject
\listrefs
\vfill\eject
\vbox{\vskip-1truecm\tabskip=0pt \offinterlineskip
      \def\tablerule{\noalign{\hrule}}
      \halign to 295pt{\strut#&\vrule#\tabskip=1.em plus2em
                   &#\hfil &\vrule#
                   &\hfil#&\vrule#
                   &\hfil#&\vrule#
                  &\hfil#&\vrule#\tabskip=0pt\cr\tablerule
             &&\omit\hidewidth parm. \hidewidth
             &&\omit\hidewidth AB\hidewidth
             &&\omit\hidewidth AR\hidewidth
             &&\omit\hidewidth OS \hidewidth&\cr\tablerule
&& $\eta_q$          && $0.48\pm 0.09$  && $0.42\pm 0.07$  && $0.35\pm 0.04$
&\cr
&& $\eta_g$          && $1.52\pm 0.74$  && $1.11\pm 0.52$  && $0.99\pm 0.23$
&\cr
&& $\alpha_{\rm NS}$ && $-0.68\pm 0.15$ && $-0.67\pm 0.13$ && $-0.70\pm 0.15$
&\cr
&& $\alpha_q$        && $0.41\pm 0.38$  && $0.34\pm 0.59$  && $0.97\pm 0.65$
&\cr
&& $\alpha_g$        && $-0.47\pm 0.30$ && $0.03\pm 0.63$  && $0.35\pm 0.67$
&\cr
&& $\beta_{\rm NS}$  && $2.2\pm 0.3$    && $1.3\pm 0.3$    && $1.3\pm 0.3$
&\cr
&& $\beta_q$         && $3.3\pm 1.4$   && $2.1\pm 1.0$    && $0.8\pm 1.2$
&\cr
&& $\beta_g$         && $2.6\pm 4.8$    && $7.0\pm 8.3$    && $4$ (fixed)
&\cr
&& $a_{\rm NS}$      && $16\pm 17$      && $8.6\pm 8.2$        && $15\pm 16$
&\cr
&& $a_q=a_g$         && $0.1\pm 3.0 $   && $0.9\pm 4.4$    && $-1.25\pm 0.09$
&\cr
\tablerule
&& $\chi^2$         &&  $62.6/63$   &&        $61.0/63$    &&       $61.8/64$
&\cr
\tablerule
}}\hfil\medskip
\centerline{\vbox{\hsize= 380pt \raggedright\noindent\footnotefont
Table~1: Best-fit values of the parameters eq.~\parm\ and $\chi^2$
for fits in the AB, AR and OS
schemes;
the errors shown are statistical only.
}}
\bigskip
\bigskip
\vbox{\tabskip=0pt \offinterlineskip
      \def\tablerule{\noalign{\hrule}}
      \halign to 415pt{\strut#&\vrule#\tabskip=0.5em 
                   &#\hfil&\vrule#
                   &\hfil#&\vrule#
                   &\hfil#&\vrule#
                   &\hfil#&\vrule#
                   &\hfil#&\vrule#
                   &\hfil#&\vrule#
                   &\hfil#&\vrule#\tabskip=0pt\cr\tablerule
       &&\omit&&\omit\hidewidth $\eta_q$\hidewidth
             &&\omit\hidewidth $\eta_g$\hidewidth
             &&\omit\hidewidth $\Gamma_1^p(10)$\hidewidth
             &&\omit\hidewidth $\Gamma_1^d(10)$\hidewidth
             &&\omit\hidewidth $a_0(10)$\hidewidth
             &&\omit\hidewidth $\chi^2$\hidewidth&\cr\tablerule
&& as tab.~1 &&$0.48\pm  0.09$   && $1.52\pm 0.74$
            &&$0.122\pm 0.013$ &&$0.025\pm 0.013$ && $0.14\pm 0.10$  && $62.6$
&\cr\tablerule
&& high $b_{\rm NS}$&&$0.48\pm 0.06$    && $1.43\pm 0.48$
             && $0.124\pm 0.011$ &&$0.027\pm0.011$  && $0.15\pm 0.08$  &&
$63.2$
&\cr
&& low $b_{\rm NS}$&&$0.50\pm0.10$     && $1.55\pm 0.81$
             && $0.123\pm 0.015$ &&$0.027\pm0.015$ && $0.15\pm 0.11$  && $64.3$
&\cr\tablerule
&& $\alpha_{\rm NS}=0$
                   &&$0.37\pm  0.09$   && $1.17\pm 0.70$
           && $0.120\pm 0.010$ &&$0.023\pm0.010$   && $0.11\pm 0.08$  && $73.9$
&\cr\tablerule
&&  high $g_A$   &&$0.48\pm  0.09$   && $1.54\pm 0.74$
           && $0.124\pm 0.013$ &&$0.025\pm0.013$  && $0.14\pm 0.10$  && $62.5$
&\cr
&& low $g_A $ &&$0.48\pm  0.09$   && $1.50\pm 0.73$
         && $0.120\pm 0.012$ &&$0.026\pm0.012$ && $0.14\pm 0.10$  && $62.7$
&\cr
&& high $a_8$    &&$0.45\pm  0.09$   && $1.58\pm 0.74$
       && $0.122\pm 0.013$ &&$0.026\pm0.013$ && $0.10\pm 0.10$  && $62.7$
&\cr
&& low $a_8$    &&$0.51\pm  0.09$   && $1.47\pm 0.73$
          && $0.122\pm 0.012$ &&$0.025\pm0.012$  && $0.18\pm 0.10$  && $62.5$
&\cr\tablerule
&& high $\alpha_s$  &&$0.51\pm  0.08$   && $1.36\pm 0.55$
           && $0.118\pm 0.013$ &&$0.022\pm0.013$  && $0.12\pm 0.10$  && $63.6$
&\cr
&& low $\alpha_s$    &&$0.47\pm  0.09$   && $1.80\pm 0.82$
           && $0.124\pm 0.011$ &&$0.026\pm0.011$  && $0.14\pm 0.09$  && $62.1$
&\cr\tablerule
&& high fact.  &&$0.42\pm  0.05$   && $1.17\pm 0.45$
           && $0.121\pm 0.010$ &&$0.023\pm0.010$ && $0.11\pm 0.08$  && $61.1$
&\cr
&& low fact.    &&$0.41\pm  0.07$   && $0.81\pm 0.39$
           && $0.133\pm 0.007$ &&$0.037\pm0.007$ && $0.25\pm 0.06$  && $63.0$
&\cr\tablerule
&& high ren. &&$0.42\pm  0.05$   && $1.31\pm 0.57$
       && $0.125\pm 0.010$ &&$0.028\pm0.010$ && $0.15\pm 0.07$  && $61.4$
&\cr
&& low ren.    &&$0.42\pm  0.08$   && $0.92\pm 0.59$
         && $0.129\pm 0.016$ &&$0.033\pm0.016$ && $0.20\pm 0.13$  && $62.3$
&\cr\tablerule
&&  high thr. &&$0.57\pm  0.06$   && $1.69\pm 0.33$
        && $0.121\pm 0.006$ &&$0.024\pm0.006$  && $0.13\pm 0.06$  && $62.5$
&\cr
&& low thr.     &&$0.46\pm  0.07$   && $1.54\pm 0.67$
          && $0.121\pm 0.012$  &&$0.025\pm0.012$ && $0.13\pm 0.10$  && $62.0$
&\cr\tablerule
}}
\hfil\medskip
\centerline{\vbox{\hsize= 380pt \raggedright\noindent\footnotefont
Table~2: Values of the parameters $\eta_q$ and $\eta_g$,
the first moment of $g_1$ eq.~\gfm, the
axial charge eq.~\axfmom\ and $\chi^2$ (64 degrees of freedom
for the entry $\alpha_{\rm NS}=0$ and 63 d.f. for all other entries),
all calculated in AB scheme,
for the various fits discussed in the text.
}}
\vfill
\eject
\supereject
\vbox{\tabskip=0pt \offinterlineskip
      \def\tablerule{\noalign{\hrule}}
      \halign to 282pt{\strut#&\vrule#\tabskip=1em plus2em
                  &\hfil#\hfil&\vrule#
                  &\hfil#\hfil&\vrule#
                  &\hfil#\hfil&\vrule#
                   &\hfil#&\vrule#\tabskip=0pt\cr\tablerule
       &&source &&\omit\hidewidth $\Delta \Gamma_1^p (10)$\hidewidth
        &&\omit\hidewidth $\Delta \Gamma_1^d (10)$\hidewidth
             &&\omit\hidewidth $\Delta a_0(10)$\hidewidth&\cr\tablerule
   && fit (statistical)     && $\pm{0.013}$ && $\pm 0.013 $
&& $\pm{0.10}$  &\cr
   && fitting procedure     && $\pm{0.003}$ && $\pm{0.003} $
&& $\pm{0.03}$  &\cr \tablerule
   && SU(2) violation       && $\pm{0.002}$ && $ \pm 0.000$
&& $\pm{0.00}$  &\cr
   && SU(3) violation       && $\pm{0.000}$ && $\pm 0.000 $
&& $\pm{0.04}$  &\cr\tablerule
   && value of $\alpha_s$   && $\epm{0.002}{0.004}$ && $\epm {0.001}{0.003}$
&& $\epm{0.00}{0.02}$
&\cr
\tablerule
   && thresholds            && $\pm 0.001$ && $ \pm 0.001$
&& $\pm 0.01 $  &\cr
\tablerule
   && higher order corrns.   && $\epm{0.011}{0.002}$ && $\epm {0.012} {0.003}$
&& $\epm{0.11}{0.03}$
&\cr
\tablerule
}}\hfil\medskip
\centerline{\vbox{\hsize= 380pt \raggedright\noindent\footnotefont
Table~3: Errors in the determination of $\Gamma_1$ and $a_0$.
}}
\vfill
\eject
\supereject
\nopagenumbers
\vsize=27truecm
\null
\vskip -4.truecm
\epsfxsize=14truecm
\hfil\epsfbox{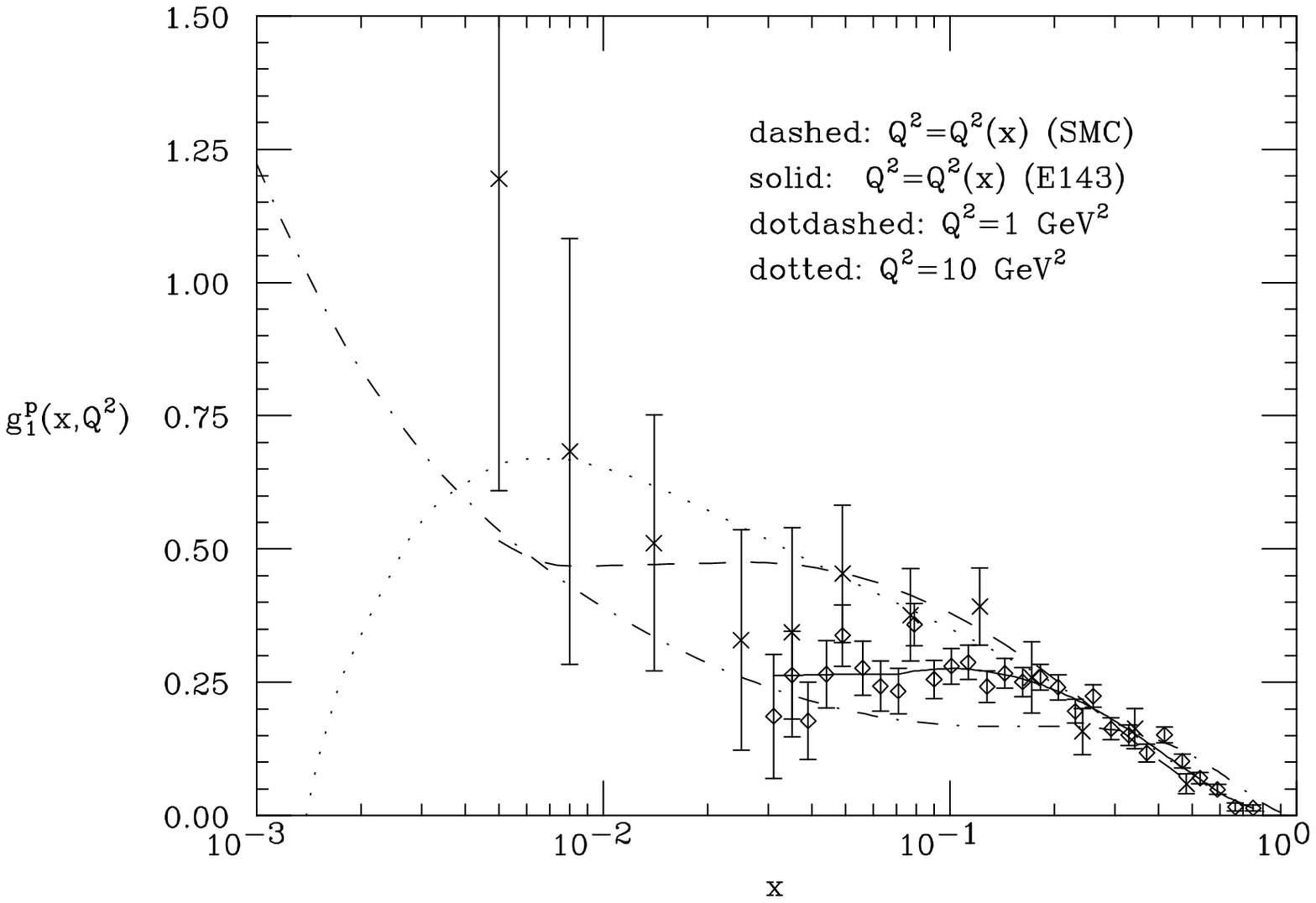}\hfil
\smallskip
\vskip -8.5truecm
\epsfxsize=14truecm
\hfil\epsfbox{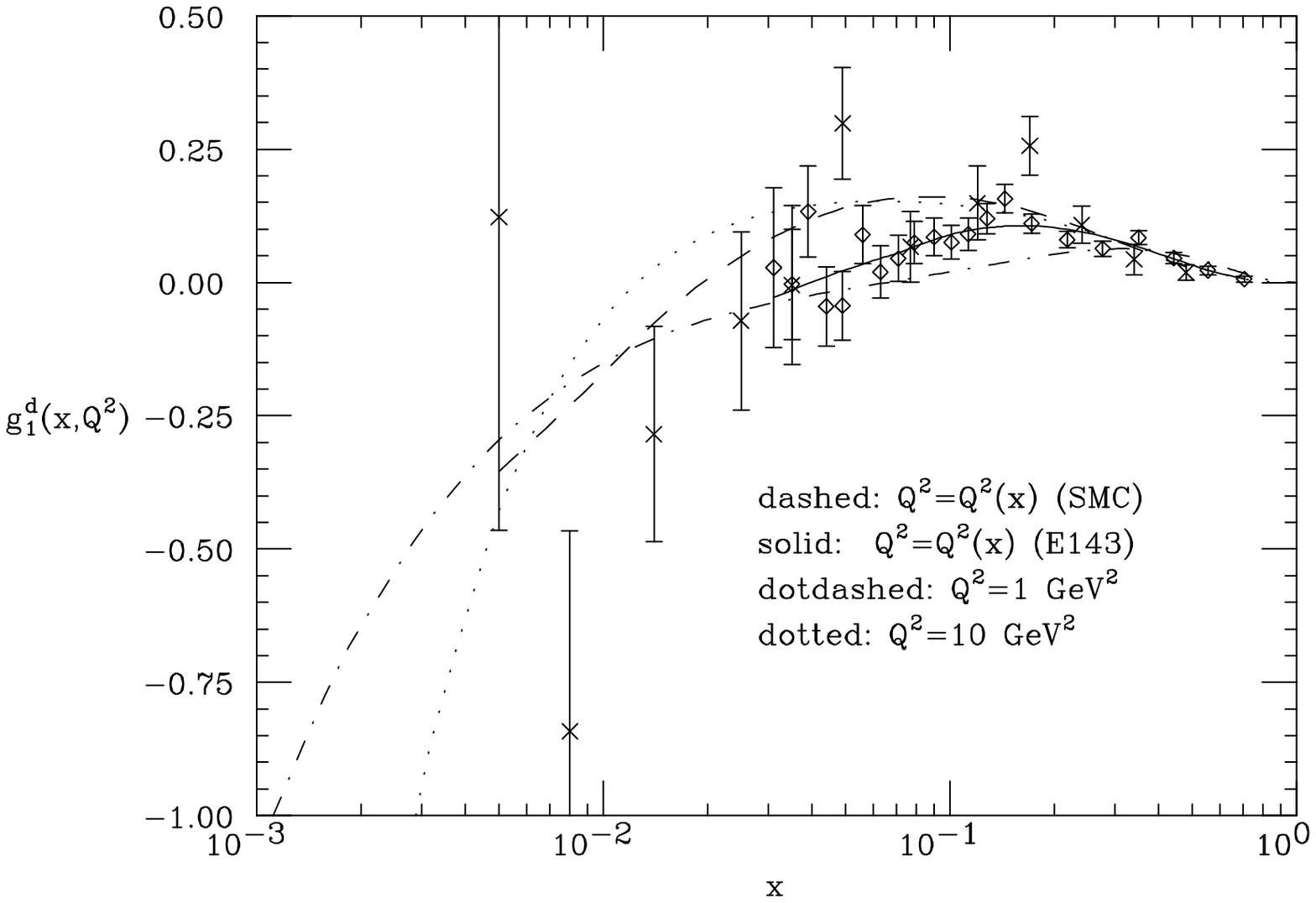}\hfil
\vskip -5truecm
\centerline{\qquad Fig.~1i}
\vfill
\eject
\null
\vskip -4.truecm
\epsfxsize=14truecm
\hfil\epsfbox{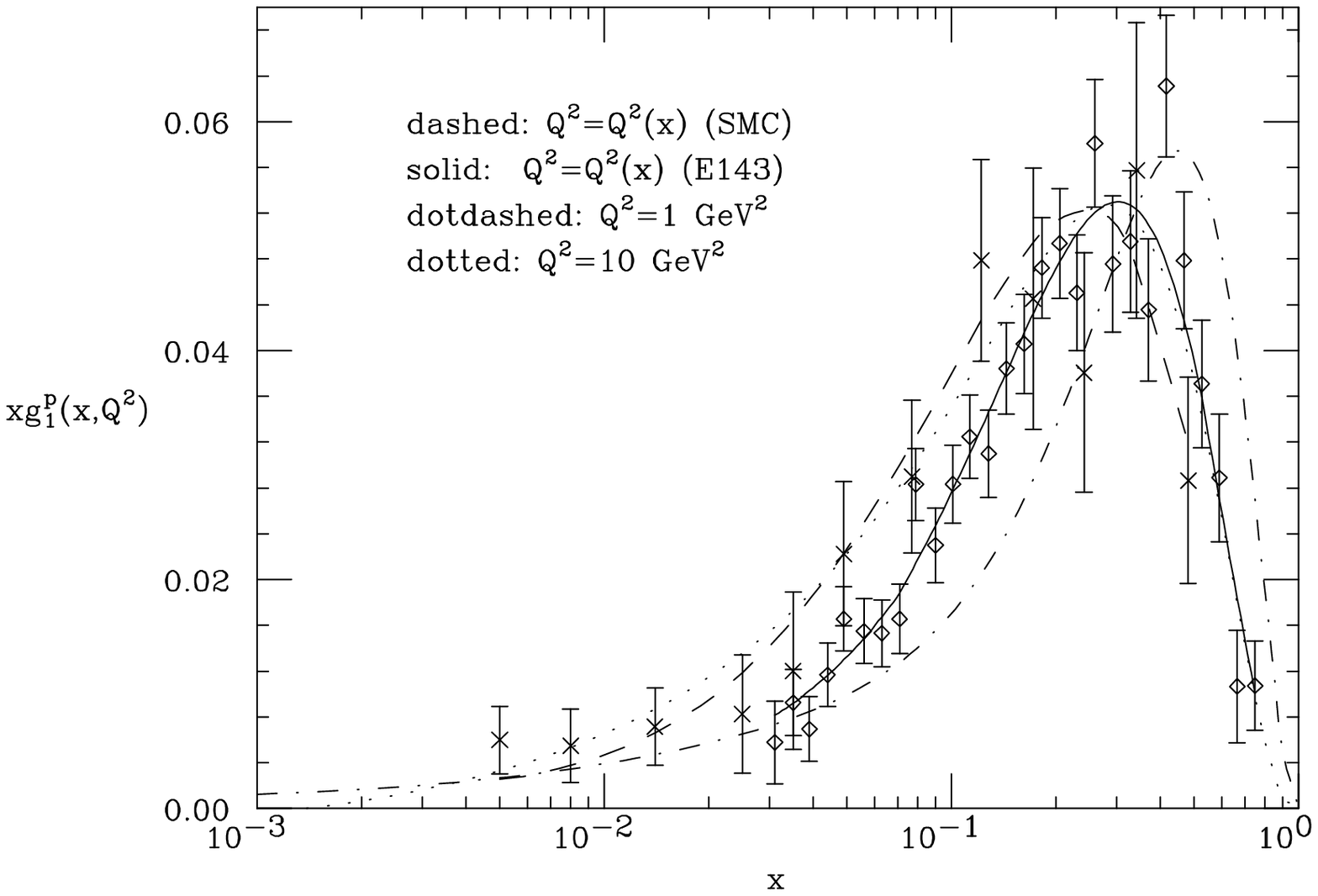}\hfil
\smallskip
\vskip -8.5truecm
\epsfxsize=14truecm
\hfil\epsfbox{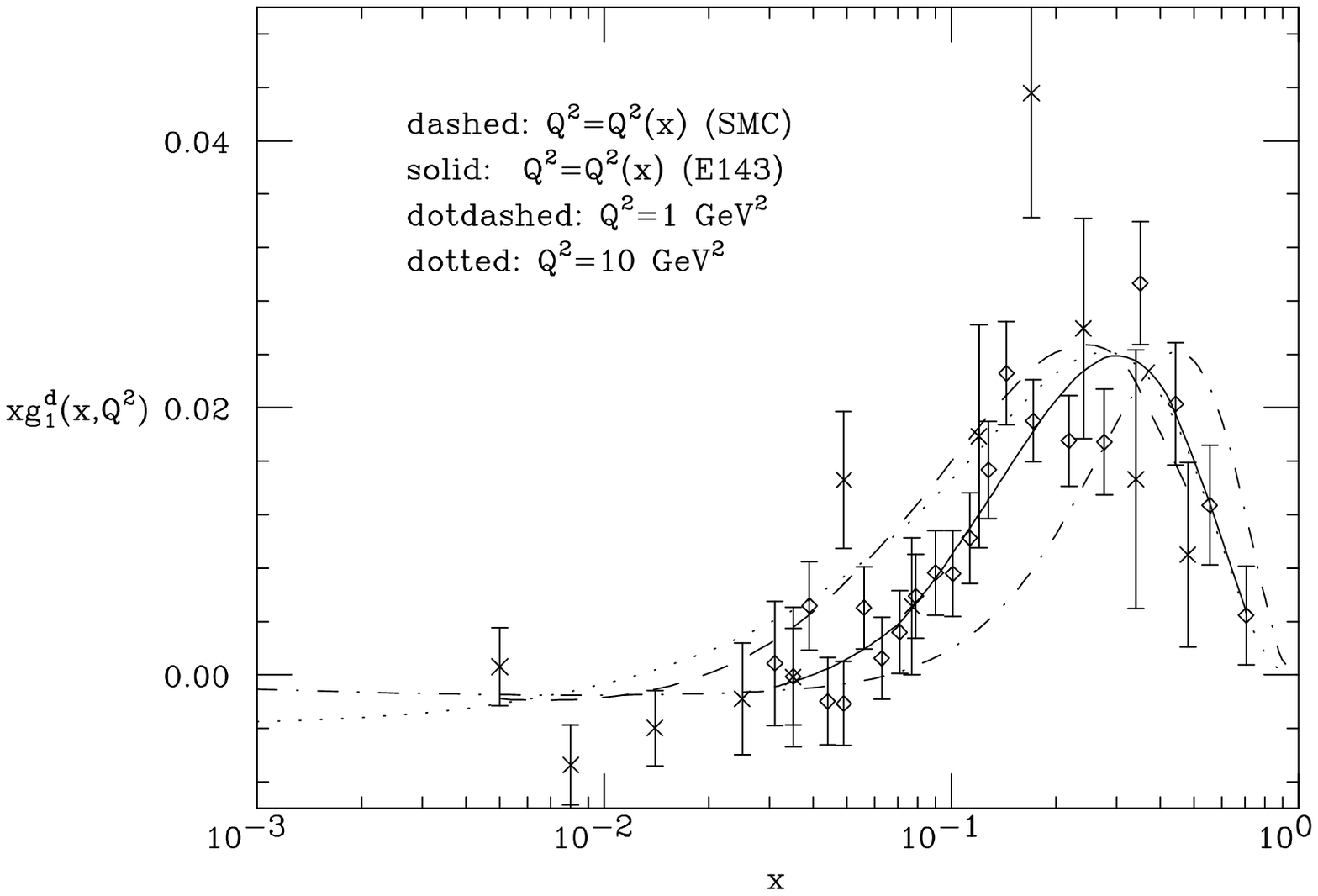}\hfil
\vskip -5truecm
\centerline{\qquad Fig.~1ii}
{\narrower \footnotefont
Plots of  $g_1(x)$ (i) and $xg_1(x)$ (ii) compared to the
SMC (crosses) and E143 (diamonds)
experimental
data for (a) proton and (b)  deuteron. The curves correspond to a NLO
computation in the AB scheme with the initial parton distributions
eq.~\parm\ and the values of the parameters given in table~1. \smallskip}
\vfill
\eject
\bye